\documentclass[aps,twocolumn,prl,superscriptaddress,footinbib]{revtex4-2}

\usepackage[usenames,dvipsnames]{color}
\usepackage[colorlinks, citecolor={blue}, urlcolor={blue}, linkcolor={blue}]{hyperref}

\usepackage{physics} 
\usepackage{amsmath} 
\usepackage{amssymb} 
\usepackage{bm} 
\usepackage{braket} 
\usepackage{latexsym} 

\renewcommand{\t}{\text} 
\newcommand{\f}[2]{\dfrac{#1}{#2}} 

\usepackage{graphicx} 
\usepackage{stackengine} 
\usepackage[caption=false]{subfig} 
\usepackage[inline]{enumitem} 
\setlist[enumerate,1]{label={(\roman*)}} 
\setlist{nolistsep} 
\usepackage{comment} 
\usepackage{lmodern} 

\usepackage{multirow}
\usepackage{hhline}

\begin{document}

\graphicspath{{./Figures/}}

\newcommand{\JILA}{JILA, National Institute of Standards and Technology and
  Department of Physics, University of Colorado, Boulder, Colorado, 80309, USA}
\newcommand{\CTQM}{Center for Theory of Quantum Matter, University of Colorado, Boulder, Colorado, 80309, USA}
\newcommand{\UCI}{Department of Physics and Astronomy, University of California, Irvine, California 92697, USA}
\newcommand{\SIT}{Department of Physics and Center for Quantum Science and Engineering, Stevens Institute of Technology, 1 Castle Point Terrace, Hoboken, New Jersey 07030, USA}
\newcommand{\Univie}{University of Vienna, Faculty of Physics, Boltzmanngasse 5, 1090 Wien, Austria}

\newcommand{\thetitle}{Validating phase-space methods with tensor networks in two-dimensional spin models with power-law interactions}

\title{\thetitle}

\author{Sean R.~Muleady}\thanks{These two authors contributed equally}
\affiliation{\JILA}
\affiliation{\CTQM}
\author{Mingru Yang}\thanks{These two authors contributed equally}
\affiliation{\UCI}
\affiliation{\Univie}
\author{Steven R.~White}
\affiliation{\UCI}
\author{Ana Maria Rey}
\affiliation{\JILA}
\affiliation{\CTQM}
\date{\today}

\begin{abstract} 

Using a recently developed extension of the time-dependent variational principle for matrix product states, we evaluate the dynamics of 2D power-law interacting XXZ models, implementable in a variety of state-of-the-art experimental platforms. We compute the spin squeezing as a measure of correlations in the system, and compare to semiclassical phase-space calculations utilizing the discrete truncated Wigner approximation (DTWA). We find the latter efficiently and accurately captures the scaling of entanglement with system size in these systems, despite the comparatively resource-intensive tensor network representation of the dynamics. We also compare the steady-state behavior of DTWA to thermal ensemble calculations with tensor networks. Our results open a way to benchmark dynamical calculations for two-dimensional quantum systems, and allow us to rigorously validate recent predictions for the generation of scalable entangled resources for metrology in these systems.

\end{abstract}

\maketitle


The understanding of how quantum correlations develop and propagate during time evolution in interacting many-body systems is a fundamental requirement for next-generation quantum technologies. While considerable work has been devoted to studying such behavior in short-ranged systems, the relatively recent experimental realization of controllable spin systems exhibiting long-ranged interactions, e.g. trapped ions~\cite{britton_engineered_2012,jurcevic_quasiparticle_2014,bruzewicz_trapped-ion_2019}, polar molecules~\cite{bohn_cold_2017,moses_new_2017}, magnetic dipoles~\cite{chomaz_dipolar_2023}, and Rydberg atoms~\cite{browaeys_many-body_2020}, has increasingly galvanized efforts to explore and characterize their utility as a quantum resource.

Despite these great opportunities, progress has been slow largely due to the lack of theoretical and numerical tools suited to faithfully characterize long-range interactions, especially in higher dimensions. The exponential growth of the size of Hilbert space with particle number typically excludes exact solutions beyond a few dozen spins. Perturbative techniques are generally restricted to short times~\cite{lepoutre_out--equilibrium_2019} or near equilibrium systems, while quasi-exact methods based on tensor network techniques have traditionally been limited to short-ranged, one-dimensional systems~\cite{schollwock_density-matrix_2011}. A host of approximate methods have been developed for this purpose, with varying ranges of applicability, including clusterization methods~\cite{tang_short_2013,hazzard_quantum_2014}, variational ansatzes~\cite{comparin_robust_2022,menu_gaussian-state_2023}, and efficiently-solved phase space methods~\cite{polkovnikov_phase_2010,schachenmayer_many-body_2015,pucci_simulation_2016,wurtz_cluster_2018,zhu_generalized_2019}, but the involved approximations are typically uncontrolled, and thus ultimately remain to be validated by other theoretical techniques or experiments.

Here, we demonstrate the utility of both tensor network methods and approximate discrete phase-space methods for studying the far-from-equilibrium collective dynamics of two-dimensional (2D) spin models with a varying range of interactions. We use a recently developed tensor network scheme based on the time-dependent variational principle (TDVP)~\cite{yang_time-dependent_2020} to solve for the collective spin dynamics of the 2D XXZ model exhibiting power-law decaying interactions. We demonstrate that the discrete truncated Wigner approximation (DTWA)~\cite{schachenmayer_many-body_2015,zhu_generalized_2019} efficiently and accurately captures the collective spin dynamics and buildup of entanglement across a wide range of parameter space, in many cases even yielding improving agreement for larger system sizes. While the large dynamical growth of entanglement eventually limits the system sizes and time accessible by TDVP, DTWA accurately captures the many-body dynamics at comparatively no computational cost, paving the avenue for its reliable use in theoretical calculations for experimentally relevant system sizes and timescales.

\begin{figure*}
 \centering
 \includegraphics[width=1.0\textwidth]{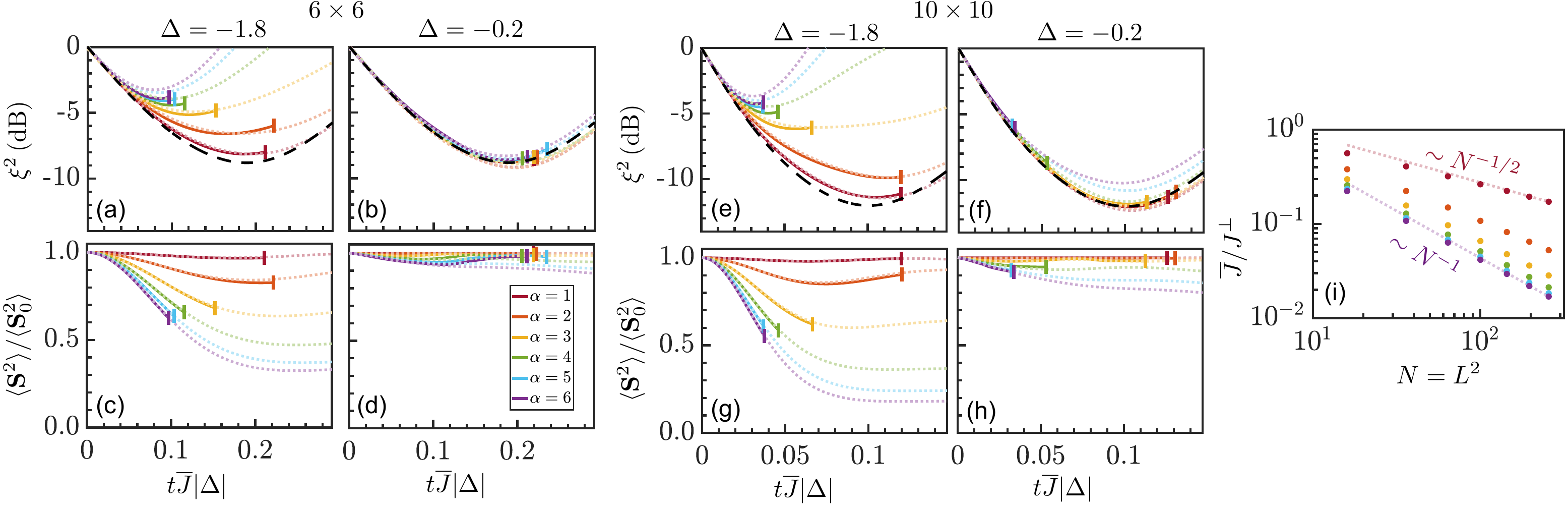}
 \caption{Dynamics of (a,b) the spin squeezing $\xi^2$ (shown in decibels) and (c,d) total spin $\langle\bm{\hat{S}}^2\rangle$ generated by Eq.~\eqref{eq:H_XXZ}, shown for select interaction ranges $\alpha$ and $\Delta = -1.8,-0.2$ on a $6\times6$ lattice. The total spin is normalized by its initial value, $\langle\bm{\hat{S}}^2\rangle_0 = (N/2)(N/2 + 1)$, and the time axis is scaled by the average spin interaction $\overline{J}$ and $\abs{\Delta}$. Solid lines are obtained by GSE-TDVP (longest evolved times denoted with vertical bars for visibility), while the dashed lines are obtained by DTWA. We show exact results for the collective case with $\alpha = 0$ for comparison (black, dashed). (e-h) Analogous results for a $10\times 10$ lattice. (i) We also plot $\overline{J}/J^{\perp}$ for various 2D square lattices of side length $L$, with expected power laws shown for comparison.}
 \label{fig:Fig_dynamics}
\end{figure*}

We further investigate the scrambling and relaxation behavior, computing the entanglement dynamics and expected thermalization temperature for this system. Though DTWA appears to capture the initial relaxation for all the investigated models, it diverges from exact calculations at long timescales for sufficiently local systems. This observation indicates the breakdown of this method for capturing thermalization in short-range interacting models. Nevertheless, our results validate the utility of DTWA for studying far-from-equilibrium dynamics of long-range, higher-dimensional spin models.

\indent{\it{Model.---}} 
We consider a system of $N$ spin-1/2 particles, with dynamics governed by the 2D power-law interacting XXZ Hamiltonian
\begin{equation}
 \hat{H}_{\t{XXZ}} = -J^{\perp} \sum_{i < j}^N \f{\hat{\bm{s}}_i\cdot \hat{\bm{s}}_j + \Delta \hat{s}_{z,i}\hat{s}_{z,j}}{|\bm{r}_i - \bm{r}_j|^{\alpha}}. \label{eq:H_XXZ}
\end{equation}
$\hat{s}_{\mu,i}$ with $\mu = x$, $y$, $z$ are spin-1/2 operators for the spin at position $\bm{r}_i$, and we assume a square lattice with open boundary conditions. We also define collective spin operators $\hat{S}_{\mu} = \sum_{i=1}^N \hat{s}_{\mu,i}$, and total spin $\hat{\mathbf{S}}^2 = \sum_{\mu=x,y,z} \hat{S}_{\mu}^2$. The Hamiltonian consists of spin-aligning terms $\hat{\bm{s}}_i\cdot \hat{\bm{s}}_j$, as well as Ising interactions $\hat{s}_{z,i}\hat{s}_{z,j}$ of relative strength $\Delta$. This canonical model of quantum magnetism with power law couplings plays a key role in describing the relevant physics for many quantum simulation platforms, including trapped ion arrays ($0\leq \alpha \leq 3$)~\cite{bruzewicz_trapped-ion_2019}, Rydberg atoms ($\alpha = 3$, $6$)~\cite{browaeys_many-body_2020}, magnetic dipoles~\cite{chomaz_dipolar_2023} and polar molecules ($\alpha = 3$)~\cite{bohn_cold_2017}, and arrays of neutral atoms ($\alpha = \infty$)~\cite{ozawa_hybrid_2021}.

We consider the spin dynamics under Eq.~\eqref{eq:H_XXZ} for an initial coherent spin state $\ket{\psi_0} = \ket{\rightarrow...\rightarrow}$, consisting of all spins polarized along $+x$. In the case of all-to-all interactions with $\alpha = 0$ or in the limit $\Delta = 0$, Eq.~\eqref{eq:H_XXZ} conserves the total spin and the state remains in the collective manifold of permutation-symmetric states. For $\alpha = 0$, the dynamics of our initial state can be described by the canonical, fully collective one-axis-twisting (OAT) model, $\hat{H}_{\rm{OAT}} = \chi\hat{S}_z^2$~\cite{kitagawa_squeezed_1993}. Additionally, in the Ising limits of $\Delta \rightarrow \pm \infty$, the local magnetization is  conserved, enabling an analytic solution for the dynamics of arbitrary two-body correlators~\cite{foss-feig_dynamical_2013}. However, away from these solvable limits, the dynamics generically involve the larger space of non-collective states.

\begin{figure*}
 \centering
 \includegraphics[width=0.99\textwidth]{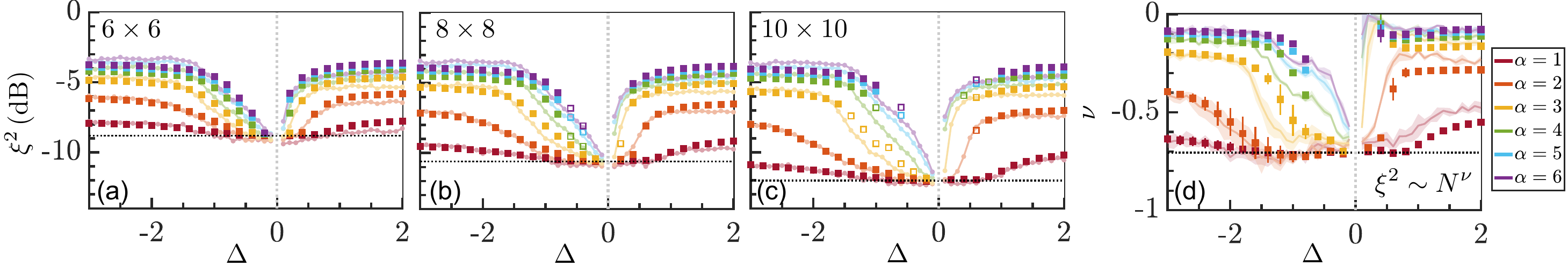}
 \caption{(a-c) Optimal spin squeezing $\xi^2$ (in decibels) computed via GSE-TDVP (squares) and DTWA (faded lines) for various 2D lattice sizes, interaction ranges $\alpha$, and $\Delta$. The dotted horizontal lines denote the expected results for the OAT model ($\alpha = 0$), while the vertical dashed lines correspond to $\Delta = 0$. Unfilled squares denote estimated values for parameters where the GSE-TDVP dynamics approach close to, but do not achieve, a local minimum for the longest evolved times~\cite{supp}. (d) Fits for the power law scaling of the optimal spin squeezing with the particle number $N$, where $\xi^2 \sim N^{\nu}$. The associated fitting error is denoted by error bars for exact results, or the shaded region for DTWA results. We supplement our data with exact/DTWA results for a $4\times 4$ lattice, and only provide fits when data for three or more system sizes are available.} \label{fig:Fig_opt}
\end{figure*}

\indent{\it{Methods.---}} 
An efficient approximation to the dynamics of Eq.~\eqref{eq:H_XXZ} is afforded by the discrete truncated Wigner approximation (DTWA)~\cite{schachenmayer_many-body_2015,schachenmayer_dynamics_2015,zhu_generalized_2019}, which provides a way to simulate the dynamics of large, interacting spin systems with a complexity polynomial in $N$. This semiclassical phase space method relies on the classical evolution of independent phase space trajectories, whose initial conditions are sampled in such a way as to exactly reproduce, within sampling error, the quantum noise distribution of an initial product state~\cite{supp}. This method, and closely related variations, have been increasingly utilized in recent years for studies ranging from the universal properties of quantum systems~\cite{babadi_far--equilibrium_2015,schuckert_nonlocal_2020} to predictions for entanglement-based sensing protocols~\cite{perlin_spin_2020,bilitewski_dynamical_2021,young_enhancing_2023,block_universal_2023}. Ultimately however, the involved approximations remain strictly valid only at short times, and the resulting predictions remain to be verified by alternative methods or experiments~\cite{schachenmayer_dynamics_2015,kunimi_performance_2021}.

As a more controlled yet intensive approach to solving the quantum dynamics, one can employ the time-dependent variational principle (TDVP)~\cite{kramer_time-dependent_1981,haegeman_time-dependent_2011,haegeman_unifying_2016}, which time-evolves a matrix product state (MPS)~\cite{schollwock_density-matrix_2011} by variationally minimizing the distance $\|\hat{H}|\psi \rangle-\mathrm{i}\frac{\mathrm{d}}{\mathrm{d}t}|\psi\rangle\|$ at each time within the tangent space of the MPS $\ket{\psi}$. Compared to other MPS time-evolution methods in the presence of long-range interactions, TDVP can retain a smaller time step error~\cite{yang_time-dependent_2020} but may accumulate large errors from projecting into the tangent space of a compressed MPS, whose number of variational parameters may be insufficient to accurately describe the evolution. To resolve this issue, we utilize a global subspace expansion (GSE) method~\cite{yang_time-dependent_2020}, which extends the bond basis of the MPS at the current-time with MPS representations of state vectors in the order-$k$ Krylov subspace,
\begin{equation}
\mathcal{K}_k(\hat{H},|\psi\rangle)=\mathrm{span}\{|\psi\rangle,\hat{H}|\psi\rangle,\dots,\hat{H}^k|\psi\rangle\}.
\end{equation}
This basis expansion enables us to more efficiently and accurately capture the relevant, developing correlations between spatially distant spins in Eq.~\eqref{eq:H_XXZ}. Nonetheless, as with any MPS method, it remains challenging to reliably evolve to long times owing to the fast bond dimension growth resulting from entanglement generation, particularly in higher spatial dimensions~\cite{supp}.

\indent{\it{Collective spin dynamics.---}} 
To examine the development of collective correlations in the dynamics, including the symmetrized off-diagonal correlators, we study the Wineland spin squeezing parameter ~\cite{wineland_spin_1992,wineland_squeezed_1994,ma_quantum_2011}, defined as
\begin{equation}
 \xi^2 \equiv \t{min}_{\hat{\bm{n}}} \f{N\langle(\Delta\hat{S}_{\hat{\bm{n}}})^2\rangle}{|\langle \hat{\bm{S}}\rangle|^2}.
\end{equation}
Here, $\langle(\Delta \hat{S}_{\hat{\bm{n}}})^2\rangle = \langle ( \hat{S}_{\hat{\bm{n}}} - \langle \hat{S}_{\hat{\bm{n}}}\rangle)^2\rangle$ is the transverse spin variance, with $\hat{S}_{\hat{\bm{n}}} = \hat{\bm{n}}\cdot \hat{\bm{S}}$, minimized over all axes $\hat{\bm{n}}$ perpendicular to the collective Bloch vector $\braket{\hat{\bm{S}}}$. This serves as a witness for multi-partite entanglement~\cite{sorensen_entanglement_2001}, and the scaling of the achievable spin squeezing with system size provides an indicator of collective behavior~\cite{comparin_robust_2022,block_universal_2023}. In addition, the spin squeezing quantifies the gain in angular resolution compared to the one from an uncorrelated coherent spin state. The sensitivity of the latter is bounded by the so-called standard quantum or shot-noise limit, given by $\xi^2_{\rm{SQL}} = 1$. A state with $\xi^2 < 1$ is thus a potential resource for improved precision measurements~\cite{pezze_quantum_2018}.

In Fig.~\ref{fig:Fig_dynamics}, we compare the performance of DTWA and GSE-TDVP in computing the spin squeezing dynamics for select values of $\Delta$ and the power-law coupling $\alpha$ for a $6\times 6$ and a $10\times 10$ system. Given the generic difficulty of simulating large system sizes, we only time-evolve with GSE-TDVP until a local minimum in the spin squeezing is reached, when possible. We also rescale the time axis by $|\Delta| \overline{J}$, where $\overline{J} = J^{\perp}\sum_{i\neq j} |\bm{r}_i - \bm{r}_j|^{-\alpha}/(N(N-1))$ is the average interaction over all spin pairs. While this provides a convenient collapse of the dynamics at early times across our parameter space, we emphasize that ``short'' timescales under this rescaling generally do not correspond to short times in terms of the underlying Hamiltonian parameters, as can be observed from the magnitude of $\overline{J}$ in Fig.~\ref{fig:Fig_dynamics}(i).

Overall, we find that DTWA excellently captures the spin squeezing dynamics. While small numerical discrepancies $\lesssim 0.5$ dB are observed in the minimum squeezing for larger $\alpha$ when $\Delta = -1.8$, this offset appears to be constant for both system sizes in consideration. Furthermore, for smaller $\alpha$ where we generally observe even better agreement, we find that any lingering discrepancies are somewhat smaller for the $10\times 10$ lattice, suggesting an improvement in accuracy as the size of the system is increased even beyond the reach of GSE-TDVP. In Fig.~\ref{fig:Fig_dynamics}(c,d,g,h) we also compare the dynamics of the total collective spin length, $\braket{\hat{\mathbf{S}}^2}$, normalized by its initial value $\langle \hat{\mathbf{S}}^2_0\rangle = \frac{N}{2}(\frac{N}{2}+1)$, finding the dynamics are well-captured by DTWA for all parameters and simulated times.

To make systematic comparisons over a wider swath of parameter space for Eq.~\eqref{eq:H_XXZ}, we utilize the optimal value of the spin squeezing (minimized over $t$) as a figure of merit for the performance of DTWA relative to GSE-TDVP. In Fig.~\ref{fig:Fig_opt}(a-c), we plot the minimum squeezing over a range of $\Delta$ and $\alpha$ for lattice sizes up to $10\times10$. We continue to find that the general agreement of DTWA with GSE-TDVP persists across all parameters considered, including a slight improvement in agreement for small $\alpha$ as the system size increases and a small, constant offset for larger $\alpha$ and $|\Delta|$. For the region $-2\lesssim \Delta < 0$, relatively longer times are required to reach the minimum squeezing compared to other $\Delta$, and we also encounter a much larger bond dimension in this regime, particularly as $\alpha$ is increased~\cite{supp}. Owing to the resource-intensive TDVP calculations for large bond dimension, we only provide results when available.

Given the demonstrated correspondence between the spin squeezing dynamics for DTWA and GSE-TDVP, as well as the rare availability of exact results for a range of 2D system sizes beyond the capabilities of exact diagonalization, we examine the ability of DTWA to capture finite-size scaling trends. Of key importance for assessing the collective nature of the dynamics, as well as the achievable utility of these states for metrology, is the exponent $\nu$ governing the system size dependence of the spin squeezing parameter~\cite{foss-feig_entanglement_2016,comparin_robust_2022,block_universal_2023}, $\xi^2 \sim N^{\nu}$. In Fig.~\ref{fig:Fig_opt}(d), we plot the values of this exponent as obtained from either DTWA or exact results for the displayed system sizes. We find good correspondence between these methods, which both capture the emergence of a quasi-collective regime with enhanced $|\nu|$ for $\alpha < 4$ and small negative $\Delta$.

\indent{\it{Thermalization.---}} To explore connections between the collective dynamics and the equilibrium physics of the XXZ model, we compare the long-time relaxation dynamics of Eq.~\eqref{eq:H_XXZ} to the thermal ensemble at the same mean energy. Given the generic difficulty of both simulating long-time dynamics and accessing the low-temperature physics of the system, we employ a combination of phase space and MPS methods to explore this regime.

We can apply DTWA to efficiently compute the long-time dynamics of this system, though owing to the large growth of entanglement as the system approaches equilibrium, we are unable to make comparisons with equivalent MPS methods, even with the GSE variant of TDVP. We thus resort to comparisons with the corresponding thermal ensemble, computed via various MPS techniques: purification~\cite{feiguin_time-step_2005} and minimally entangled typical thermal states (METTS)~\cite{white_minimally_2009,stoudenmire_minimally_2010}. Owing to the global conservation of $\hat{S}_z$ by Eq.~\eqref{eq:H_XXZ}, we include an associated Lagrange multiplier in our thermal ensemble that properly accounts for local fluctuations of this quantity, and which should be adequate to describe the thermalized values of local observables. However, for the consideration of \emph{global} correlators of the closed system, where $\hat{S}_z$ is not allowed to fluctuate, we further modify our ensemble with an additional Lagrange multiplier to enforce conservation of the variance $(\Delta\hat{S}_z)^2$~\cite{supp}.

\begin{figure}
 \centering
 \includegraphics[width=0.48\textwidth]{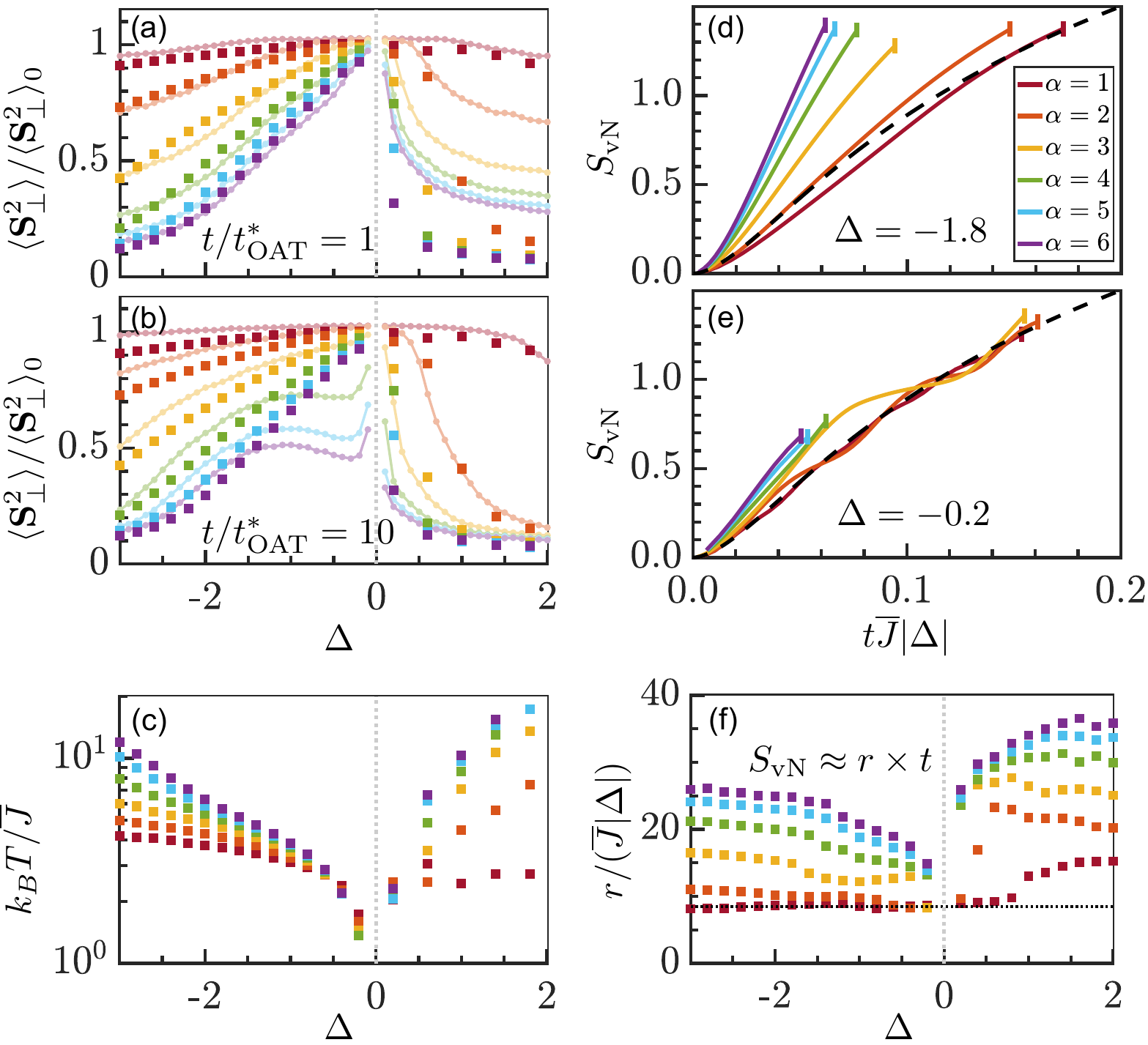}
 \caption{(a,b) Comparison between the long-time values (lines) and thermal ensemble averages (squares, same in each panel) of the transverse magnetization $\langle\hat{\mathbf{S}}_\perp^2\rangle$, normalized by its initial value of $\langle\hat{\mathbf{S}}_\perp^2\rangle_0 = N(N+1)/4$ for a $6\times 6$ lattice. Dynamical results are obtained via DTWA for various scaled times $t/t_{\mathrm{OAT}}^*$. The thermal averages are obtained via METTS for $\Delta > 0$ and via a purification method for $\Delta < 0$. (c) We also show the temperature $T$ of the thermal state, scaled by $\overline{J}$. (d/e) Dynamical growth of the bipartite entanglement entropy, $S_{\mathrm{vN}} = -\Tr\left[\hat{\rho}_A\ln \hat{\rho}_A\right]$, where $\hat{\rho}_A$ is the reduced density matrix for a bipartition of the lattice about the center. We show GSE-TDVP results on an $8\times 8$ lattice for various values of $\Delta$ and $\alpha$, scaling the time axis by $\overline{J}|\Delta|$. We also provide the results for the OAT model (black, dashed) for comparison. (f) From the available short-time data, we estimate the growth rate of the entropy by fitting $S_{\mathrm{vN}} \approx r \times t$, and plot the resulting coefficient $r$, scaled by $\overline{J}|\Delta|$. The resulting fit for the OAT model over the range $t\overline{J}|\Delta| \lesssim 0.3$ is also shown (black, dotted).}
 \label{fig:Fig_thermal}
 \label{fig:Fig_entropy}
\end{figure}

In Fig.~\ref{fig:Fig_thermal}, we plot the value of the transverse magnetization $\langle \hat{\mathbf{S}}^2_\perp\rangle$ for various scaled times as computed via DTWA, where $\hat{\mathbf{S}}_\perp = (\hat{S}_x,\hat{S}_y,0)$, and compare this to the value in the associated thermal ensemble, obtained from MPS. We find that $\braket{\hat{\mathbf{S}}_\perp^2}$ at the rescaled time $t/t_{OAT}^* = 1$, where $t_{\mathrm{OAT}}^*$ is the optimal squeezing time for an OAT model with coupling $\chi = \overline{J}|\Delta|/2$, appears to align closely with the thermal value when $\Delta < 0$. However, at later times, we observe that the value of $\braket{\hat{\mathbf{S}}_\perp^2}$, as computed via DTWA, continues to slowly decay for $\alpha > 2$ and $-2\lesssim \Delta < 0$, significantly deviating from the thermal expectation.

We attribute this artifact to a breakdown of DTWA at long times when thermalizing to the expected long-range/quasi-long-range ordered state --- the XY-ferromagnet characterized by a macroscopic transverse magnetization --- for $\alpha > 2$ and small negative $\Delta$, as opposed to any physical effect present in the actual system. Indeed, as $\Delta$ approaches $0$, the initial spin-polarized state tends to an eigenstate of the system, and should not undergo any spurious relaxation. Overall however, we find that DTWA at least plateaus to the expected thermal value initially, before exhibiting this further decay. For $\Delta > 0$ and $\Delta \lesssim -2$, the relaxation at late times of DTWA agrees well with METTS and purification calculations.

In Fig.~\ref{fig:Fig_thermal}(c), we also plot the temperature of the thermal state obtained via purification/METTS, which we scale by $\overline{J}$ to account for the possible super-extensive scaling of Eq.~\eqref{eq:H_XXZ}. We can see that the onset of a large total spin in the late-time dynamics and corresponding thermal ensemble is linked closely with a reduction in temperature, as well as an enhanced scalability of the attainable spin squeezing in Fig.~\ref{fig:Fig_opt}(d), suggesting the continuous-symmetry-broken low energy structure plays a critical role in the onset of collective squeezing behavior~\cite{perlin_spin_2020,comparin_robust_2022,block_universal_2023}.

Finally, in Fig.~\ref{fig:Fig_entropy}(d-f), we estimate the growth of the bipartite entanglement entropy over the 2D lattice from available GSE-TDVP dynamics, which is a quantity that remains inaccessible to DTWA. Close to $\Delta = 0$, we find that the entropy growth rate is similar for all $\alpha$, and collapses close to the expected entropy growth for the all-to-all case in our scaled time. This is consistent with the scrambling of the dynamics within the limited set of collective states. Away from $\Delta = 0$, we observe faster entanglement growth for shorter-ranged interactions, where the dynamics spreads beyond the collective manifold to the larger space of non-collective states. This regime coincides with an observed decrease in the attainable spin squeezing, as well as with an increase in the size of the bond dimensions required to represent the state~\cite{supp}. Overall, we observe that an increased growth rate appears to be roughly associated with a higher final temperature of the corresponding thermal ensemble.

\indent{\it{Conclusions.---}} 
We have demonstrated that DTWA provides efficient, accurate solutions for the dynamics of the 2D XXZ model with power-law decaying interactions through benchmarks with an extended TDVP algorithm, on lattice sizes of up to $10\times 10$. These system sizes are directly relevant for current experiments utilizing trapped ions~\cite{franke_quantum-enhanced_2023} or optical tweezer arrays of Rydberg atoms~\cite{bornet_scalable_2023,eckner_realizing_2023}. We have also probed the long-time relaxation of DTWA with MPS calculations of the expected thermal ensemble, finding good agreement at intermediate times and the ability to capture the onset of long-range/quasi-long-range order, while also demonstrating the continued utility of MPS methods for probing inaccessible quantities to DTWA, such as the entanglement growth.

\begin{acknowledgments}
\noindent{\it Acknowledgements:} We acknowledge helpful discussions with Chunlei Qu and Michael Perlin, as well as valuable feedback on the manuscript from David Wellnitz and Alexey Gorshkov. The MPS calculations were performed on the Vienna Scientific Cluster (VSC) using the ITensor~\cite{fishman_itensor_2022} and ITensor/TDVP~\cite{yang_time-dependent_2020} library. AMR and SRM are  supported by the AFOSR Grant No. FA9550-18-1-0319, the ARO single investigator Grant No. W911NF-19-1-0210, the NSF PHY1820885, NSF JILA-PFC PHY-1734006 and NSF QLCI-2016244 grants, by the DOE Quantum Systems Accelerator (QSA) grant and by NIST.
MY has received support from the European Research Council (ERC) under the European Union's Horizon 2020 research and innovation program through the ERC Consolidator Grant SEQUAM (Grant No. 863476). SRW acknowledges the support of the NSF through grant DMR-2110041.
\end{acknowledgments}

\bibliography{LRSpins20,supp}

\newpage

\onecolumngrid
\vspace{\columnsep}
\begin{center}
\textbf{\large Supplemental Material for ``Validating phase-space methods with tensor networks in two-dimensional spin models with power-law interactions''}
\end{center}
\vspace{\columnsep}

\setcounter{equation}{0}
\setcounter{figure}{0}
\setcounter{table}{0}
\setcounter{page}{1}
\makeatletter
\renewcommand{\theequation}{S\arabic{equation}}
\renewcommand{\thefigure}{S\arabic{figure}}

\section{Discrete truncated Wigner approximation}
The discrete truncated Wigner approximation (DTWA)~\cite{polkovnikov_phase_2010,schachenmayer_many-body_2015,zhu_generalized_2019} provides an efficient way to simulate the coherent dynamics of large, interacting spin systems. This semiclassical phase space method relies on the classical evolution of independent phase space trajectories, whose initial conditions are sampled in such a way so as to exactly reproduce (within sampling error) the initial quantum noise distribution of an initial product state. We represent the spin-1/2 operators for the $i$-th spin by a set of classical phase-space variables, i.e. $\hat{\bm{s}}_i = (\hat{s}_{x,i},\hat{s}_{y,i},\hat{s}_{z,i}) \longrightarrow \bm{\mathcal{S}}_i = (\mathcal{S}_{x,i},\mathcal{S}_{y,i},\mathcal{S}_{z,i})$, and impose canonical angular momentum Poisson bracket relations $\{\mathcal{S}_{\mu,i},\mathcal{S}_{\nu,j}\} = -\delta_{ij}\sum_{\rho}\varepsilon_{\mu\nu\rho}\mathcal{S}_{\rho,i}$ for Levi-Civita symbol $\varepsilon_{\mu\nu\rho}$, analogous to the spin commutation relations via $i[\hat{s}_{\mu,i},\hat{s}_{\nu,j}] \longrightarrow \{\mathcal{S}_{\mu,i},\mathcal{S}_{\nu,j}\}$. We can then obtain the classical equations of motion for these variables
\begin{align}
    \frac{d}{dt}\mathcal{S}_{\mu,i} = \left\{\mathcal{H},\mathcal{S}_{\mu,i}\right\},\label{eq:H_classical}
\end{align}
where the Weyl symbol $\mathcal{H}$ for our Hamiltonian is obtained from $\hat{H}$ by a straightforward replacement of the spin operators with their corresponding phase space variables.

To describe the state of the system, we introduce a Wigner quasiprobability distribution $W(\{\bm{\mathcal{S}_i}\})$ over the phase space variables, where $\int d\bm{\mathcal{S}}_1 ... d\bm{\mathcal{S}}_N W(\left\{\bm{\mathcal{S}}_i\right\}) = 1$. While various representations of states have been found to yield predictions of varying accuracy for specific models~\cite{schachenmayer_many-body_2015,pucci_simulation_2016,kunimi_performance_2021}, here we utilize the generic scheme outlined in Ref.~\cite{zhu_generalized_2019}, which is straightforward to implement across models can be readily generalized to systems with larger local Hilbert spaces~\cite{lepoutre_out--equilibrium_2019,morong_disorder-controlled_2021}. For a single spin initially pointing along $+x$, $\ket{\rightarrow}$, the associated initial Wigner distribution may be written
\begin{equation}
\begin{split}
    W(\bm{\mathcal{S}}_i) = \frac{1}{4}\delta(\mathcal{S}_{x,i} - 1/2)\times\left[\delta(\mathcal{S}_{y,i} - 1/2) + \delta(\mathcal{S}_{y,i} + 1/2)\right]\times\left[\delta(\mathcal{S}_{z,i} - 1/2) + \delta(\mathcal{S}_{z,i} + 1/2)\right].
    \end{split}
\end{equation}
The classical evolution of this distribution can be simulated by implementing a Monte Carlo sampling of initial conditions from this distribution, and independently evolving the resulting trajectories under the classical evolution equations in Eq.~\eqref{eq:H_classical}. This amounts to each trajectory being initialized in one of the four configurations
\begin{align}
    (\mathcal{S}_{x,i},\mathcal{S}_{y,i},\mathcal{S}_{z,i}) \in \left\{\left(\frac{1}{2},\frac{1}{2}, \frac{1}{2}\right), \left(\frac{1}{2},\frac{1}{2}, -\frac{1}{2}\right), \left(\frac{1}{2},-\frac{1}{2}, \frac{1}{2}\right), \left(\frac{1}{2},-\frac{1}{2}, -\frac{1}{2}\right)\right\},
\end{align}
each with probability $1/4$. For many spins initialized in a spin-polarized state along $+x$, each set of spin variables may be independently sampled according to the above distribution. Here, we use $10^4$ trajectories for our calculations.

At any given time $t$, the Wigner distribution can be approximated by the distribution of the classical trajectories. To compute spin observables and symmetrically ordered correlators, one can average the corresponding variables over the initial phase space distribution, i.e.
\begin{equation}
    \langle\hat{s}_{\mu,i}\rangle \approx \int d\bm{\mathcal{S}}_1 ... d\bm{\mathcal{S}}_N W(\left\{\bm{\mathcal{S}}_i\right\}) \mathcal{S}_{\mu,i} = \overline{\mathcal{S}_{\mu,i}}\label{eq:avg_1}
\end{equation}
\begin{equation}
    \langle\hat{s}_{\mu,i}\hat{s}_{\nu,j}\rangle \approx \int d\bm{\mathcal{S}}_1 ... d\bm{\mathcal{S}}_N W(\left\{\bm{\mathcal{S}}_i\right\}) \mathcal{S}_{\mu,i}\mathcal{S}_{\nu,j} = \overline{\mathcal{S}_{\mu,i}\mathcal{S}_{\nu,j}},\label{eq:avg_2}
\end{equation}
where $\overline{\,\,\cdot\,\,}$ denotes averaging over our Monte Carlo trajectories. 

In computing on-site correlators, $\braket{\hat{s}_{\mu,i}\hat{s}_{\nu,i}}$, there is an ambiguity in how to obtain these from DTWA. We may either 1) reduce the spin correlator to a single-body observable, i.e. $\hat{s}_{\mu,i}\hat{s}_{\nu,i} = \varepsilon_{\mu\nu\rho}\hat{s}_{\rho,i}/2 + \delta_{\mu\nu}/4$, and then approximate this quantity using Eq.~\eqref{eq:avg_1}, or 2) directly approximate the correlator via $\overline{\mathcal{S}_{\mu,i}\mathcal{S}_{\nu,i}}$ in Eq.~\eqref{eq:avg_2}. In practice, these two procedures yield different results: when $\mu = \nu$, the correlator should be a constant, but computing this via the second procedure leads to unexpected dynamics. Nonetheless, as found here and in related studies~\cite{schachenmayer_many-body_2015,pucci_simulation_2016,zhu_generalized_2019,perlin_spin_2020,young_enhancing_2023}, in the computation of collective quantities this latter method generally yields improved results, and we thus adopt this approach. One reason for seeing why this may be the preferred method is that the quantity $\braket{\hat{S}_z^2}$ should be conserved under the XXZ Hamiltonian we consider in the main text. Likewise, the associated classical equations of motion conserve the value of $\sum_i\mathcal{S}_{z,i}$ for each trajectory, and likewise the value of $\sum_{i,j}\mathcal{S}_{z,i}\mathcal{S}_{z,j}$ will be conserved. However, there is no symmetry in the classical equations guaranteeing the conservation of $\sum_{i\neq j}\mathcal{S}_{z,i}\mathcal{S}_{z,j} + N/4$, which represents the value of $\hat{S}_z^2$ when we reduce on-site two-body observables to single-body observables. Thus, this latter quantity will not be representative of the expected dynamics (or lack thereof) for the collective $z$ correlator.

The number of variables in DTWA is linear in $N$, yielding an efficient simulation of the dynamics that is polynomial in $N$. Although we utilize the classical equations of motion, we emphasize that the nonlinear equations acting on our noise distribution can lead to the development of non-trivial correlations with time, e.g. $\overline{\mathcal{S}_{\mu,i}\mathcal{S}_{\nu,j}} \neq \overline{\mathcal{S}_{\mu,i}}\cdot\overline{\mathcal{S}_{\nu,j}}$. This generally results dynamics that far more accurately capture many aspects of the quantum dynamics compared to classical mean-field theory. Ultimately, however, the involved approximations remain strictly valid only at short times \cite{polkovnikov_phase_2010}, though varying degrees of agreement with the full quantum dynamics beyond these times has been observed in a number of settings~\cite{schachenmayer_dynamics_2015,pucci_simulation_2016,lepoutre_out--equilibrium_2019,kunimi_performance_2021,alaoui_measuring_2022}.

\section{Time-dependent variational principle with global subspace expansion}
We employ the time-dependent variational principle (TDVP)~\cite{kramer_time-dependent_1981,haegeman_time-dependent_2011,haegeman_unifying_2016}, which is a time evolution method for matrix product states (MPS)~\cite{schollwock_density-matrix_2011}. This method involves projecting the time-dependent Schr{\"o}dinger equation to the tangent space of the variational manifold $\mathcal{M}$ of the MPS $|\psi [A]\rangle$ at the current time $t$, i.e.
\begin{equation}
\mathrm{i}\frac{\mathrm{d}}{\mathrm{d}t}|\psi [A]\rangle=\hat{P}_{T\mathcal{M}}\hat{H}|\psi [A]\rangle,
\end{equation}
where $\hat{P}_{T\mathcal{M}}$ is the projector to the one-site or two-site tangent space of the MPS $|\psi [A]\rangle$. After a Lie-Trotter decomposition, we can then solve the resulting local effective equations site-by-site, and thus variationally optimize the MPS to minimize the distance $\|\hat{H}|\psi [A]\rangle-\mathrm{i}\frac{\mathrm{d}}{\mathrm{d}t}|\psi [A]\rangle\|$.

\begin{figure}[h!]
 \centering
 \includegraphics[width=\textwidth]{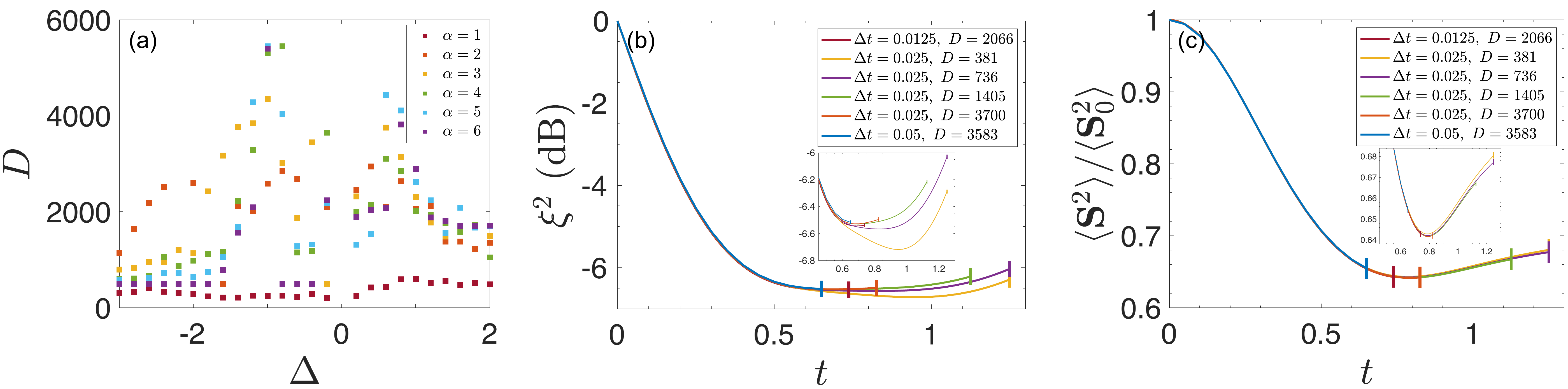}
 \caption{(a) The maximum bond dimension reached by TDVP for the real-time dynamics of the $10\times 10$ systems. Near $\Delta=0$ for $3\leq\alpha\leq 6$, we note the minimum spin squeezing time, $t^*$, is unattainable in our simulations. (b) The spin squeezing parameter $\xi^2$ in simulations with different time steps $\Delta t$ and maximum bond dimensions $D$ for the $10\times 10$ lattice with $\alpha=3$ and $\Delta=-1.6$. (c) Dynamics of $\langle\hat{\mathbf{S}}^2\rangle$, scaled by its initial $\langle\hat{\mathbf{S}}^2\rangle_0 = (N/2)(N/2+1)$, in simulations with varying time steps $\Delta t$ and maximum bond dimensions $D$ for the $10\times 10$ system at $\alpha=3$ and $\Delta=-1.6$. The insets in (b) and (c) provide a close-up of the curves near their local minima. The truncation error of the MPS per unit time in different simulations is fixed to be $\epsilon=10^{-8}$ per $\Delta t=0.025$, until the maximum bond dimension $D$ is reached. For this plot, we have set $J^{\perp} = 1$.}
 \label{fig:converge}
\end{figure}

As noted in the main text, TDVP has a natural advantage for dealing with long-range interactions while retaining small time step errors compared to other MPS time-evolution methods~\cite{yang_time-dependent_2020}. However, TDVP suffers from pitfalls due to the potentially large errors from projecting into the tangent space of a finite-bond-dimension MPS, whose number of variational parameters may be far from sufficient to accurately describe the resulting state, especially for long-range interacting systems. To resolve this issue, a global subspace expansion (GSE) method~\cite{yang_time-dependent_2020} was proposed to enlarge the tangent space so that the projection errors are reduced. Specifically, before each TDVP time step, the method enriches the bond basis of the MPS at the current time by adding important basis vectors from the compressed MPS representation of state vectors in the order-$k$ Krylov subspace,
\begin{equation}
\mathcal{K}_k(\hat{H},|\psi\rangle)=\mathrm{span}\{|\psi\rangle,\hat{H}|\psi\rangle,\dots,\hat{H}^k|\psi\rangle\}.
\end{equation}
Through this approach, we are allowed to use larger time steps while maintaining precision. When the bond dimension eventually grows to a computationally unwieldy size in the time evolution, we can switch from the GSE extension over to the conventional two-site TDVP to continue our calculations to even longer times.

This basis expansion enables us to efficiently capture the relevant, developing correlations between spatially distant spins in the system. Nonetheless, as with any MPS method, the main obstacle to performing real-time dynamics is the fast bond dimension growth resulting from entanglement, as shown in Fig.~\ref{fig:converge}(a). This challenge is compounded by the large bond dimension ($\sim 100$) of the matrix product operator for the power-law interacting Hamiltonian~\footnote{In principle, one could compress the MPO by approximating the power-law with a finite sum of exponentials, as in Ref.~\cite{zaletel_time-evolving_2015}, for use with other time-evolution methods such as time-evolving block decimation. However, our use of TDVP alleviates the need for this approach.}. Thus, it remains challenging to reliably evolve to long times, and we only evolve long enough to observe an extremum in the spin squeezing, when possible. When we are unable to fully evolve to this time, denoted $t^*$, we can apply an order-7 polynomial fitting and Fourier fitting to predict the subsequent evolution $\xi^2$ and $\braket{\hat{\mathbf{S}}^2}$, respectively. In Fig.~\ref{fig:Fig_opt} of the main text, we show the results of these predictions when the spin squeezing appears to be sufficiently close to reaching a local minimum based on the longest evolved time, and denote these to distinguish from simulations that have reached $t^*$.

With a time step $\Delta t=t^*/24$, we first apply one-site TDVP with GSE. We utilize $k=3$, with truncation cutoffs $\epsilon_K=10^{-8}$ for $\hat{H}|\psi\rangle$ and $\epsilon_K=10^{-6}$ for $\hat{H}^2|\psi\rangle$; in addition, we utilize a truncation cutoff $\epsilon_M=10^{-4}$ for controlling the number of added orthogonal basis vectors from the Krylov subspace, and a truncation cutoff $\epsilon=10^{-10}$ during the TDVP step~\footnote{Our notation for the GSE parameters matches Ref.~\cite{yang_time-dependent_2020}.}. We evolve to a time $t^*/6$, and then switch to ordinary two-site TDVP with truncation cutoff $\epsilon=10^{-8}$ for each time step. By comparing to test simulations using the GSE time evolution for all time steps, we find that the error induced by switching to ordinary two-site TDVP evolution is negligible.

In Fig.~\ref{fig:converge}(b,c), we show the results from various TDVP simulations with different time steps $\Delta t$ and maximum bond dimensions $D$ for the $10\times 10$ system at $\alpha=3$ and $\Delta=-1.6$, and find that using a time step $\Delta t=0.025/J^{\perp} \approx t^*/24$ and $D=3174$ is sufficient to achieve convergence in the value of $t^*$ and the corresponding minimum value of $\xi^2$, as well as in the value of $\langle\hat{\mathbf{S}}^2\rangle$ at $t^*$. From these plots, we also note that 1) decreasing $D$ appears to induce a larger minimum in the spin squeezing $\xi^2$ as well as increase the value of $t^*$, and 2) $\langle\hat{\mathbf{S}}^2\rangle$ converges faster with increasing bond dimension than $\xi^2$.

\section{Purification and minimally-entangled typical thermal states}
Owing to the rapid growth of entanglement in the dynamics, evolving to long times to observe a steady state is generally extremely difficult via MPS techniques. Alternatively, we can apply equilibrium MPS methods to examine the thermal state of the system, and study connections to the associated steady state dynamics computed efficiently via DTWA, where the system presumably thermalizes. Computing thermal properties at finite temperatures in MPS requires translating thermal ensembles to pure states, or otherwise involves imaginary time evolution of a matrix product density operator~\cite{verstraete_matrix_2004}. In this work, we utilize two different strategies --- purification~\cite{feiguin_time-step_2005} and minimally entangled typical thermal states (METTS)~\cite{white_minimally_2009,stoudenmire_minimally_2010} --- to obtain thermal properties at $\Delta < 0$ and $\Delta > 0$, respectively.

The purification of a mixed state $\hat{\rho}=\sum_{a=1}^rs_a^2|a\rangle_P\langle a|_P$ of a physical system $P$ is given by
\begin{equation}
|\psi\rangle=\sum_{a=1}^r s_a|a\rangle_P|a\rangle_Q,
\end{equation}
where the ancilla $Q$ is a replica of $P$. Thus, $\hat{\rho}$ can be understood as the partial trace of the density matrix of a pure state $|\psi\rangle$ over $Q$, i.e. $\mathrm{Tr}_Q|\psi\rangle\langle\psi |$.

A canonical ensemble at $T = 1/\beta$ can then be written as
\begin{equation}
\hat{\rho}_\beta=\frac{\mathrm{Tr}_Q|\mathrm{TFD}\rangle\langle\mathrm{TFD}|}{\langle\mathrm{TFD}|\mathrm{TFD}\rangle},
\end{equation}
where $|\mathrm{TFD}\rangle$ is the thermofield double state
\begin{equation}
|\mathrm{TFD}\rangle=\frac{1}{\sqrt{Z(\beta)}}e^{-\beta\hat{H}/2}|\psi_0\rangle,
\end{equation}
and $|\psi_0\rangle$ is the purification of the mixed state at $\beta=0$, which maximally entangles $P$ and $Q$. The expectation value of an observable $\hat{O}$ is thus given by
\begin{equation}
\langle \hat{O}\rangle_\beta=\mathrm{Tr}_P\left[\hat{O}\hat{\rho}_\beta\right]=\frac{\langle\mathrm{TFD}|\hat{O}|\mathrm{TFD}\rangle}{\langle\mathrm{TFD}|\mathrm{TFD}\rangle}.
\end{equation}
In this way, the mixed state at $T=1/\beta$ is obtained by performing imaginary time evolution on a pure state $|\psi_0\rangle$ to $\beta/2$.

While we are interested in the low-temperature properties, the entanglement in the purification is approximately doubled~\cite{schollwock_density-matrix_2011} when $T\sim 0$ compared to that in the density matrix renormalization group (DMRG)~\cite{white_density_1992}, which is designed to target the pure ground state. Thus, the required bond dimension of the MPS in purification is approximately the square of that in DMRG, and the imaginary time evolution of the purification can be quite demanding.

To avoid this problem, we can alternatively use METTS. The key idea of METTS is based on Monte Carlo sampling over a carefully chosen set of states such that each can be represented efficiently by an MPS of much smaller bond dimension than in the purification. In Monte Carlo, it is standard to write $\langle\hat{O}\rangle_\beta$ as
\begin{equation}
\begin{split}
\langle\hat{O}\rangle_\beta &=\frac{1}{Z(\beta)}\mathrm{Tr}\left[\hat{O}e^{-\beta\hat{H}}\right]\\
&=\frac{1}{Z(\beta)}\sum_i\langle i|e^{-\beta\hat{H}/2}\hat{O}e^{-\beta\hat{H}/2}|i\rangle\\
&=\sum_i\frac{P_\beta(i)}{Z(\beta)}\langle \phi_\beta(i)|\hat{O}|\phi_\beta(i)\rangle\\
&=\overline{\langle \phi_\beta(i)|\hat{O}|\phi_\beta(i)\rangle},
\end{split}
\end{equation}
where $\{|i\rangle\}$ is a set of orthonormal basis states of the Hilbert space for the system. The state
\begin{equation}
|\phi_\beta(i)\rangle=P_\beta(i)^{-1/2}e^{-\beta\hat{H}/2}|i\rangle
\end{equation}
is called a METTS if $|i\rangle$ is chosen to be a classical product state, with the normalization factor $P_\beta(i)=\langle i|e^{-\beta\hat{H}}|i\rangle$. We can estimate $\langle\hat{O}\rangle_\beta$ by sampling $|\phi_\beta(i)\rangle$ with the probability $P_\beta(i)/Z(\beta)$. This probability distribution can be obtained by generating a Markov chain of states with the transition probability
\begin{equation}
T_{\beta}(i\rightarrow i')=|\langle i'|\phi_\beta(i)\rangle|^2
\end{equation}
which satisfies the detailed balance
\begin{equation}
P_{\beta}(i)T_{\beta}(i\rightarrow i')=P_{\beta}(i')T_{\beta}(i'\rightarrow i).
\end{equation}

Starting from a random classical state $i_1$, the $k$-th iteration of the algorithm proceeds as follows:
\begin{enumerate}
    \item Perform imaginary time evolution and normalization to $|i_k\rangle$ and get $|\phi_\beta(i_k)\rangle$.
    \item Evaluate $\langle \phi_\beta(i_k)|\hat{O}|\phi_\beta(i_k)\rangle$.
    \item Collapse the state $|\phi_\beta(i_k)\rangle$ to a new classical product state $|i_{k+1}\rangle$ by quantum measurements according to the transition probability $T_{\beta}(i_k\rightarrow i_{k+1})$.
\end{enumerate}
We iterate the above steps to obtain a Markov chain of length $R$. The thermal average is then approximated by
\begin{equation}
\langle\hat{O}\rangle\approx \frac{1}{R}\sum_{k=1}^R\langle \phi_\beta(i_k)|\hat{O}|\phi_\beta(i_k)\rangle.
\end{equation}

\end{document}